\documentstyle[aps]{revtex}
\newcommand{\bd}{\begin{displaymath}}
\newcommand{\ed}{\end{displaymath}}
\newcommand{\be}{\begin{equation}}
\newcommand{\ee}{\end{equation}}
\newcommand{\barr}{\begin{eqnarray}}
\newcommand{\earr}{\end{eqnarray}}
\newcommand{\barrr}{\begin{eqnarray*}}
\newcommand{\earrr}{\end{eqnarray*}}
\newcommand{\ua}{\uparrow}
\newcommand{\da}{\downarrow}

\title{Fragmentation of\\
$SU(2)$-invariant spin ladders}
\author{\em Giuseppe Albertini\thanks{E-mail 
Giuseppe.Albertini@mi.infn.it}\\
\em Dipartimento di Fisica, Universita' di Milano\\
\em via Celoria 16, 20133 Milano (Italy)} 
\date{20.2.2001}

\begin{document}

\maketitle

\begin{abstract}
A two-parameter family of quantum spin ladders with local bilinear
and biquadratic interactions is shown to be solvable by a mapping onto
fragments of integrable spin 1 chains. The phase diagram, consisting
of four phases, and the ground state properties are discussed. In
one novel phase, the ground state is made up of plaquette singlets
and rung singlets, alternating with a three-rung periodicity.
\end{abstract}

\\
\\
IFUM-681-FT 


\section{Introduction}
Integrable models have provided extremely valuable insight into the
physics of one dimensional quantum spin chains. There is a substantial
lack of exact results for analogous two dimensional systems and
the solution of quantum spin ladders might provide a first step in that
direction. Besides, spin ladders are currently experimentally
accessible, interesting in their own and possibly related to high $T_{c}$
superconductivity \cite{DR}.

It has been established that the basic $n$-leg Heisenberg ladders, with
bilinear exchange interactions along rungs and legs, are gapful (spin-liquid
state) for $n$ even and gapless for $n$ odd \cite{DR}. 
These ladders do not seem to be integrable.
On the other hand, examples of integrable ladders, containing additional
biquadratic interactions, have been found and solved by some form of
Bethe-ansatz (BA) \cite{Wa,AFW}. Luckily, biquadratic interactions do arise 
in physically realizable systems, and a large class of these generalized, 
but still $SU(2)$-invariant ladders, have been proven to have a 
matrix-product (MP) ground state \cite{KM}. 
Still, the MP approach determines the ground 
state but, with the exception of few lucky cases, the whole set of 
excitations remains unknown.

In this paper, the following two-parameter family of $SU(2)$-invariant
ladder hamiltonians $H= \sum_{k=1}^N H_{k,k+1}$ will be studied
\barr\label{e1}
&&H_{k,k+1}(\tilde{\mu}_1,\tilde{\mu}_2)=\frac{1}{2} 
({\bf s}_k \cdot {\bf s}_{k+1}+{\bf t}_k \cdot {\bf t}_{k+1}+
{\bf s}_k \cdot {\bf t}_{k+1}+{\bf s}_{k+1} \cdot {\bf t}_{k}) \nonumber \\
&&-2\Big(({\bf s}_k \cdot {\bf s}_{k+1})( {\bf t}_{k} \cdot {\bf t}_{k+1}) +
({\bf s}_k \cdot {\bf t}_{k+1})( {\bf s}_{k+1} \cdot {\bf t}_{k}) \Big) +
\tilde{\mu}_1({\bf s}_k \cdot {\bf t}_k + {\bf s}_{k+1} \cdot {\bf t}_{k+1})
\nonumber \\
&&+\tilde{\mu}_2 ({\bf s}_k \cdot {\bf t}_k)({\bf s}_{k+1} 
\cdot {\bf t}_{k+1})
\earr
and the eigenvalue problem solved for any $\tilde{\mu}_1$, $\tilde{\mu}_2$.
Here ${\bf s}_k$ and ${\bf t}_k$ are spin 1/2 matrices sitting on the first
and the second leg, respectively.
Hamiltonian (\ref{e1}) is actually a special point in a wider 
three-parameter class for which the relative strength of the first 
two terms is left arbitrarty, (see (\ref{e11})). All points in this 
manifold have the property of being reducible, through the introduction 
of the composite rung spin  
\be\label{e2}
{\bf S}_k = {\bf s}_k + {\bf t}_k
\ee
to the general spin 1 chain 
\be\label{e3}
H_{0}(\theta) = \sum_{k=1}^N \Big(\cos\theta{\bf S}_k \cdot {\bf S}_{k+1} + 
\sin\theta ({\bf S}_k \cdot {\bf S}_{k+1})^2 \Big)
\ee
In particular (\ref{e1}) corresponds to the BA solvable
purely biquadratic spin 1 chain, $\theta = -\pi/2$. Selecting 
$\theta = 0$, the Heisenberg spin 1 chain, would remove the second 
brackets in (\ref{e1}) \cite{Xi}. This seems physically 
more appealing but prevents the detailed analysis allowed by BA, 
since such chain is not integrable. Details about definitions and exact
diagonalization are the subject of sections II and III.  

The system defined by (\ref{e1}) has four phases, connected by first order 
transition lines, namely discontinuities in some first order derivative of
the ground state energy per site. At least one phase is novel because its
ground state is a threefold degenerate global singlet where dimerization
takes place alternatively along legs and rungs with a three-step
periodicity. A second phase has a huge ($\sim 3^{N/2}$) ground state
degeneracy. All this is discussed in section IV.

In section V, exact elementary excitations are examined in the phase whose
ground state is, effectively, that of the biquadratic chain, also a
global singlet but made up of rung triplets. The simplest ones are created
by introducing one rung singlet and leaving the remaining $N-1$ triplets
locked into the ground state of the biquadratic chain with free ends.
If $N-1$ is odd, such ground state is presumably an $SU(2)$ triplet 
containing one dynamical kink and
one finds a band of singlet-triplet excitations, depending on two degrees
of freedom, 
whose energy is degenerate in the singlet position. Such excitations 
had been found in a related ladder \cite{Xi}, but their dependence 
on two degrees of freedom had not
been discussed. On the other hand, similar ladders including biquadratic
interactions are known to have two-parameter singlet-triplet 
excitations \cite{NT,KM}, but their nature is different, being 
a pair of kink-antikink over a dimerized ground state, which is not 
what happens here. In section VI, some general results are discussed 
for the wider three-parameter family of spin ladder hamiltonians.

For the sake of completeness, the Appendix contains a review of known results
on the biquadratic chain.

\section{Definition of basic invariants}
Each elementary plaquette involves four spins, say $
({\bf s}_{1}, {\bf t}_{1}, {\bf s}_{2}, {\bf t}_{2})$. If only $SU(2)$
invariance is required, the most general hermitean plaquette hamiltonian
$H_{1,2}$ is a linear combination with real coefficients of 14 
hermitean invariants. If, beside $SU(2)$
invariance, one requires symmetry under the exchange of the two legs
and symmetry under the exchange of the two rungs, symmetries
implemented by operators $C$ and $P$
\barrr
&&C{\bf s}_{i}C = {\bf t}_{i} \ \ (i=1,2) \hspace{2cm} C^2=1  \\
&&P{\bf s}_{1}P = {\bf s}_{2} \hspace{1cm} P{\bf t}_{1}P = {\bf t}_{2} 
\hspace{1cm} P^2=1
\earrr
then $H_{1,2}$ is a linear combination with real coefficients of 8 
hermitean invariants. 

The proof goes as follows. The local plaquette Hilbert space is 
${\mathcal{H}}_{1,2}$ $\simeq V^{(s)}_1 \otimes V^{(t)}_1 \otimes V^{(s)}_2
\otimes V^{(t)}_2 \simeq C^{16}$. It breaks into the orthogonal sum
of $SU(2)$ multiplets: one quintuplet (dim=5), three triplets (dim=9)
and two singlets (dim=2). Within each multiplet, all states are obtained
by application of the lowering operator ${\bf S}^{-}_{1,2}$
to the highest weight vector (h.w.v.) $v^{(s)}$, where
\barrr
{\bf S}_{1,2}={\bf s}_1 + {\bf s}_2 + {\bf t}_1 + {\bf t}_2 \hspace{1cm}
{\bf S}^{\pm}_{1,2}={\bf S}^{x}_{1,2}\pm i{\bf S}^{y}_{1,2}  \\
{\bf S}_{1,2}^{+}v^{(s)}=0 \hspace{1cm} {\bf S}^2_{1,2}v^{(s)} =
s(s+1)v^{(s)} \hspace{1cm} v^{(s)} \in {\mathcal{H}}_{1,2}
\earrr
$H_{1,2}$ is $SU(2)$-invariant, i.e. $[H_{1,2},{\bf S}_{1,2}]=0$, 
therefore it maps h.w.v. into h.w.v. of the same spin.
So, spin-2 h.w.v. must be mapped into itself, each
spin-1 h.w.v. will in general be mapped into a linear combination of all
three spin-1 h.w.v. and the same happens for the two spin-0 h.w.v.. 
Altogether, imposing hermiticity, 6 real parameters plus 4 complex (hence
8 real) ones.
The total spin ${\bf S}_{1,2}$ commutes with
$C$ and $P$, so h.w.v. can be labelled by $C$ and $P$ eigenvalues. Define,
with self-explanatory notation 
\barrr
&& e_0 = |\ua\ua>_s \ \ f_0 = |\ua\ua>_t \ \ 
\overline{e}_0 = |\da\da>_s \ \ \overline{f}_0 = |\da\da>_t \\
&& e_1 = |\da\ua>_s \ \ f_1 = |\da\ua>_t \ \ e_2 = |\ua\da>_s \ \  f_2 =
|\ua\da>_t
\earrr
The 6 h.w.v. $v^{(s)}$ will be chosen to be
\barrr
v^{(2)} &=& e_0f_0 \hspace{7.3cm} (2,1,1)\\
v^{(1)}_1 &=& e_0f_1 - e_0f_2 +e_1f_0 - e_2f_0 \hspace{4cm} (1,1,-1)\\
v^{(1)}_2 &=& e_0f_1 + e_0f_2 - e_1f_0 - e_2f_0 \hspace{4cm} (1,-1,1)\\
v^{(1)}_3 &=& e_0f_1 - e_0f_2 - e_1f_0 + e_2f_0 \hspace{4cm} (1,-1,-1)\\
v^{(0)}_1 &=& e_1f_1 + e_2f_2 - e_1f_2 - e_2f_1 \hspace{4cm} (0,1,1) \\
v^{(0)}_2 &=& e_1f_1 + e_1f_2 + e_2f_1 + e_2f_2 - 2e_0\overline{f}_0
-2\overline{e}_{0}f_{0} \hspace{1.5cm} (0,1,1)
\earrr
The three numbers on the right column are eigenvalues of $S_{1,2}$, 
$C$, $P$. If $C$ and $P$ symmetries are imposed,
$H_{1,2}$ maps $v^{(2)}$ into itself and each $v^{(1)}_{i}$ $(i=1,2,3)$ into 
itself, while its action on the span of $\{v^{(0)}_i; i=1,2\}$ is 
determined by 4 real numbers
\barrr
H_{1,2}v^{(2)} &=& m^{(2)} v^{(2)} \hspace{1cm} H_{1,2}v^{(1)}_{i} =
m^{(1)}_i v^{(1)}_i \hspace{1cm} i=1,2,3 \\
H_{1,2}v^{(0)}_i &=& \sum_{j=1}^2 m_{ji}^{(0)} v_j^{(0)} \hspace{1cm} i=1,2 
\earrr
This proves that the general $H_{1,2}$ is 
a linear combination with real coefficients of 8
linearly independent invariants (notice incidentally that h.w.v. $v^{(0)}_i$
are orthogonal but unnormalized, $\Vert v^{(0)}_2 \Vert ^2 = 
3 \Vert v^{(0)}_1 \Vert ^2$, so $m_{12}^{(0)} = 3m_{21}^{(0)*}$).
One can choose 7 of them to be
three bilinear (Heisenberg) terms
\barr
I_{1,2}^{(1)} & = & {\bf s}_1 \cdot {\bf s}_2 + {\bf t}_1 \cdot {\bf t}_2 
\label{e4} \\
I_{1,2}^{(2)} & = & {\bf s}_1 \cdot {\bf t}_1 + {\bf s}_2 \cdot {\bf t}_2 
\label{e5} \\
I_{1,2}^{(3)} & = & {\bf s}_1 \cdot {\bf t}_2 + {\bf s}_2 \cdot {\bf t}_1
\label{e6} 
\earr
three biquadratic (plaquette) terms
\barr
I_{1,2}^{(4)} & = & ({\bf s}_1 \cdot {\bf s}_2)({\bf t}_1 \cdot {\bf t}_2) 
\label{e7} \\
I_{1,2}^{(5)} & = & ({\bf s}_1 \cdot {\bf t}_1)({\bf s}_2 \cdot {\bf t}_2) 
\label{e8} \\
I_{1,2}^{(6)} & = & ({\bf s}_1 \cdot {\bf t}_2)({\bf s}_2 \cdot {\bf t}_1)
\label{e9} 
\earr
and the identity $\mathcal{I}$. An eighth one is necessary to have a complete
set, but it involves more complicated combinations of the basic spins and
it will not be needed in the following.
To show that the six in (\ref{e4})-(\ref{e9}) plus the identity
are indeed linearly independent, write
\bd
H_{1,2} = c_0{\mathcal I}+\sum_{k=1}^6 c_k I_{1,2}^{(k)}
\ed
It is a matter of easy algebra to find the action of the basic invariants
(\ref{e4})-(\ref{e9}) on the h.w.v., resulting in
\barr\label{e10}
m^{(2)} &=& c_0 + \frac{c_1+c_2+c_3}{2} + \frac{c_4+c_5+c_6}{16} \nonumber  \\
m^{(1)}_1 &=& c_0 + \frac{-c_1+c_2-c_3}{2} + 
\frac{-3c_4+c_5-3c_6}{16} \nonumber \\
m^{(1)}_2 &=& c_0 + \frac{c_1-c_2-c_3}{2} + 
\frac{c_4-3c_5-3c_6}{16} \nonumber \\
m^{(1)}_3 &=& c_0 + \frac{-c_1-c_2+c_3}{2} + \frac{-3c_4-3c_5+c_6}{16} \\
m^{(0)}_{11} &=& c_0 + \frac{-3c_1}{2} + 
\frac{9c_4+3c_5+3c_6}{16} \nonumber \\  
m^{(0)}_{22} &=& c_0 + \frac{c_1-2c_2-2c_3}{2} + 
\frac{c_4+7c_5+7c_6}{16} \nonumber \\
m^{(0)}_{21} &=& \frac{c_2-c_3}{2} + \frac{-c_5+c_6}{8} \nonumber
\earr
and $m^{(0)}_{12}=3m^{(0)}_{21}$.
The vanishing of all $\{m^{(s)}_i\}$ implies the vanishing of all
$\{c_i\}$, proving linear independence of the invariants chosen.

Scalar products of the rung spin will now be expressed in terms of
these invariants. Clearly
\bd
{\bf S}_1 \cdot {\bf S}_2 = I^{(1)}_{1,2} + I^{(3)}_{1,2}
\ed
while the biquadratic term $({\bf S}_1 \cdot {\bf S}_2)^2$
seems to involve more complicated invariants. But since
it is invariant under $C$ and $P$, it must be a linear combination of the
8 basic ones. Actually, the six (\ref{e4})-(\ref{e9}) and the identity 
are sufficient: from (\ref{e10})
\bd
({\bf S}_1 \cdot {\bf S}_2)^2 = \frac{3{\mathcal I}}{4} - 
\frac{I^{(1)}_{1,2}}{2} + I^{(2)}_{1,2} - \frac{I^{(3)}_{1,2}}{2}
+2I^{(4)}_{1,2} +2I^{(6)}_{1,2}
\ed
Consider now the one-parameter spin ladder Hamiltonian (\ref{e3})
where each spin ${\bf S}$ is the composite object defined in (\ref{e2}).
Two parameters can be added after noticing that the charges
\bd
Q_1 = \sum_{k=1}^N {\bf S}_k^2 \hspace{2cm} Q_2 = \sum_{k=1}^N {\bf S}_k^2 
{\bf S}_{k+1}^2
\ed
commute with $H_{0}(\theta)$ for any $\theta$. This can easily be checked by a
direct calculation. Finally, the Hamiltonian to be studied is
\barr\label{e11}
&&H(\theta,\mu_1,\mu_2) =H_0(\theta) + \mu_1 Q_1 + \mu_2 Q_2 = \nonumber \\
&&\sum_{k=1}^N \bigg[ (\cos\theta-\frac{\sin\theta}{2})(I_{k,k+1}^{(1)}
+I_{k,k+1}^{(3)}) + 2\sin\theta (I_{k,k+1}^{(4)} +I_{k,k+1}
^{(6)}) + (\sin\theta + \mu_1 + 3\mu_2) I_{k,k+1}^{(2)} \nonumber \\  
&& + 4\mu_2 I_{k,k+1}^{(5)} \bigg]
+N(\frac{3\sin\theta}{4}+\frac{3\mu_1}{2}+\frac{9\mu_2}{4}) 
\earr
Eq. (\ref{e1}) is (\ref{e11}), up to a constant shift, if one 
takes $\theta =-\pi/2$ and
\be\label{e12}
\tilde{\mu}_1 = -1 +\mu_1 +3 \mu_2 \hspace{1cm} \tilde{\mu}_2 = 4\mu_2
\ee
In the following, parameters $(\mu_1,\mu_2)$ will be adopted to describe the
model. It is always possible, through (\ref{e12}), to revert 
to the original ones $(\tilde{\mu}_1,\tilde{\mu}_2)$.

One further remark. Since each rung space $V^{(s)}_k \otimes V^{(t)}_k$ carries
a spin $\underline{1} \oplus \underline{0}$ representation, not just
spin 1, $({\bf S}_k \cdot
{\bf S}_{k+1})^3$ is linearly independent from ${\bf S}_k \cdot {\bf S}_{k+1}$
and $({\bf S}_k \cdot {\bf S}_{k+1})^2$. But, from (\ref{e10}), 
it can be seen that
$({\bf S}_k \cdot {\bf S}_{k+1})^3$ is a linear combination of the first
two powers, ${\mathcal I}$, $I^{(2)}_{k,k+1}$ and $I^{(5)}_{k,k+1}$, 
that is why adding a cubic term to $H_0(\theta)$ does not 
generalize (\ref{e11}).

The composite rung spin has, of course, been introduced in several previous
works. It lies behind the physical idea that even-legged and odd-legged
Heisenberg ladders should be in different phases \cite{DR}. More closely to
this work, it has been used \cite{KM} to map two-leg ladders into 
the AKLT chain \cite{AKLT}
((\ref{e3}) at $\tan \theta =1/3$). Earlier, Xian and then Kitakani
and Oguchi \cite{Xi} had studied (\ref{e11}) at
$\theta =0$, but without the rung-rung interaction $I^{(5)}_{k,k+1}$
which is actually responsible for the rise of two new phases.

\section{Fragmentation and Bethe-ansatz}

Charges $Q_1$ and $Q_2$ have a simple physical meaning. In each four
dimensional rung space $V_k=V^{(s)}_k \otimes V^{(t)}_k \simeq C^4$ 
one introduces the singlet-triplet basis $\{|s>_k;|t>_k, t=-1,0,1\}$ 
consisting of the singlet and the triplet of the rung spin 
(\ref{e2}), i.e. 
\bd
{\bf S}_k^2 |s>_k=0 \hspace{1cm} {\bf S}_k^2 |t>_k = 2|t>_k 
\hspace{1cm} S_k^z |t>_k = t|t>_k 
\ed
Then $\frac{Q_1}{2}$
counts the number of triplets and $\frac{Q_2}{4}$ counts the number of
pairs of neighboring triplets.

Now consider the $3^N$ dimensional subspace (of the total $4^N$-
dimensional Hilbert space of the ladder) spanned by
$|t_1,t_2,....,t_N>$,  $(t_k=-1,0,1)$.
In this sector $Q_1$ and $Q_2$ are constant, at $2N$ and $4N$ respectively,
and $H_0$ acts effectively as a spin-1 chain with periodic boundary
conditions (p.b.c.). Such chain is Bethe-ansatz solvable in three cases:
$(a)$ $\theta=\frac{\pi}{4}$, $(b)$ $\theta=-\frac{\pi}{4}$ and $(c)$ 
$\theta=-\frac{\pi}{2}$
In case $(a)$ it is the $SU(3)$-invariant Sutherland-Uimin chain \cite{Ui},
in case $(b)$ the Babujian-Takthajian chain \cite{Ba} and in case $(c)$ the
purely biquadratic chain \cite{Kl}.

Next, consider states containing one singlet and $N-1$ triplets. Eigenvalues
of the two conserved charges are fixed at
\bd
Q_1 = 2(N-1) \hspace{2cm} Q_2 = 4(N-2)
\ed
regardless of triplet's position. If the singlet is, say, on the $N^{th}$ rung
(due to p.b.c. the singlet's position is actually immaterial), $H_0$ acts on
these vectors like
\be\label{e13}
H_0 \simeq \sum_{k=1}^{N-2} (\cos \theta{\bf S}_k \cdot {\bf S}_{k+1} + 
\sin \theta ({\bf S}_k \cdot {\bf S}_{k+1})^2)
\ee
that is exactly like a spin-1 chain of length $N-1$ and free
boundary conditions (f.b.c.). The singlet can be positioned anywhere, it just 
opens a fracture in a ring of spins, consequently each eigenvalue of 
(\ref{e13}) appears with an $N$-fold degeneracy in the ladder spectrum.
More importantly, the spectrum of (\ref{e13}) can be found exactly 
by Bethe-ansatz, at least in cases $(b)$ and $(c)$ which have been 
more or less extensively studied for f.b.c. \cite{BB,MNR}. 
The opposite situation arises when all rungs are singlets. There is only one
such state and its eigenvalue is trivially zero.

Between these extreme cases, and orthogonal to them, there is all 
possible intermediate situations, characterized by alternating fragments of
triplets and singlets. Sectors are labeled by a sequence of positive
integers, the lengths of fragments, 
$\{N^{(0)}_1, N^{(1)}_1,
N^{(0)}_2, N^{(1)}_2,..., N^{(0)}_n, N^{(1)}_n \}$, with 
$\sum_{j=1}^n N_j^{(0)} = N^{(0)}$, $\sum_{j=1}^n N_j^{(1)} =N^{(1)}$ and
$N^{(0)}+N^{(1)} = N$. Each sector is spanned by the $3^{N^{(1)}}$ vectors
\bd
|\phi> = |\underbrace{s,s,...,s}_{N^{(0)}_1};
\underbrace{t^{(1)}_1,t^{(1)}_2,...,
t^{(1)}_{N^{(1)}_1}}_{N^{(1)}_1}; \underbrace{s,s,...,s}_{N^{(0)}_2};
 \underbrace{t^{(2)}_1,t^{(2)}_2,...,t^{(2)}_{N^{(1)}_2}}_{N^{(1)}_2};...>
\ed
On all these basis vectors
\barrr
&&H_{0} |\phi> =  \sum_{j=1}^n H_0^{(j)} |\phi> \\
&&H_0^{(j)} |\phi> =  |s,s,...,s>_{N^{(0)}_1} \otimes |t^{(1)}_1,t^{(1)}_2,
...,t^{(1)}_{N^{(1)}_1}>_{N^{(1)}_1} \otimes ... \otimes H_0^{(j)} 
|t^{(j)}_1,...,t^{(j)}_{N^{(1)}_j}>_{N^{(1)}_{j}} \\ 
&&\otimes  |s,s,...,s>_{N^{(0)}_{j+1}} \otimes ...
\earrr
where $H_{0}^{(j)}$ acts on the string of triplets like a spin-1 chain of
length $N^{(1)}_j$ and f.b.c.. Call $|\psi>_{N^{(1)}_j}$ any of its 
eigenvectors and $E_j(N^{(1)}_j)$ the relevant eigenvalue. Then
\be\label{e14}
|\psi> = |s,s,...,s>_{N^{(0)}_1} \otimes |\psi_1>_{N^{(1)}_1} \otimes
|s,s,...,s>_{N^{(0)}_2} \otimes ... \otimes |\psi_n>_{N^{(1)}_n}
\ee
is an eigenvector of $H_{0}$ with eigenvalue
\be\label{e15}
E = \sum_{j=1}^n E_j(N^{(1)}_j)
\ee
Vectors in (\ref{e14}) provide a complete set for the whole ladder 
to the extent the vectors $|\psi_j>_{N_j}$ provide a complete set 
of eigenvectors for
the spin-1 chain of length $N_j$, so diagonalization of the spin 1 chain
with p.b.c. and f.b.c. provides a complete solution to the diagonalization
problem of the spin ladder (\ref{e11}).
In general this is only a partial simplification.
Instead, in cases $(a)$, $(b)$ and $(c)$ eigenvalues can
in principle be found, for all fragments, by the suitable Bethe-ansatz.
What happens here is very similar, actually almost identical, 
to what happens in the mixed Heisenberg chains studied by Niggemann
{\em et al.} \cite{NUZ}. 
There, composite rung spins alternate with single spins and, 
very much like it happens here, fragmentation takes place when one 
rung spin is in a singlet state. In the following I will partly 
adopt notation and methods introduced in their work.

\section{The ground state problem}

To identify the ladder ground state one must minimize (\ref{e15}). 
An easy first step is to choose, in (\ref{e15}), 
the lowest eigenvalue for each fragment. To fix
the notation, denote with $E^{(p)}_0(N;1)$ the lowest eigenvalue of
the $N$-site spin 1 chain with p.b.c. and with $E^{(f)}_0(N';1)$ the
lowest one for the $N'$-site chain with f.b.c.. Then, in a sector
$\{N_j^{(0)},N_j^{(1)}\}_{j=1}^n$ the lowest eigenvalue is
\be\label{e16}
E=\left\{ \begin{array}{ll}
\sum_{j=1}^n \Big( E^{(f)}_0 (N_j^{(1)};1)+(2\mu_1+4\mu_2) N^{(1)}_j - 
4\mu_2 \Big)
& N^{(1)} \neq 0,N \nonumber \\
E^{(p)}_0 +(2\mu_1+4\mu_2)N & N^{(1)}=N \\
0 & N^{(1)} = 0 \nonumber
\end{array} \right.
\ee
Following \cite{NUZ}, a spin 0 rung is joined to, say, 
the right of each triplets 
fragment, bringing its length to $j+1$. Set $n_j$ to be the number of triplet 
fragments of length $j+1$ (i.e. $j$ triplets plus one singlet
at the edge). Consider, for the time being, only fragmented configurations,
i. e. $N^{(0)} > 0$,
specified by $\{n_0,n_1,n_2, \ldots \}$ which must satisfy
\bd
\sum_{j=0}^{N-1} (j+1) n_j = N
\ed
Their energy is 0 for $n_0=N$, otherwise
\bd
E(n_0,n_1, \ldots, n_N) = \sum_{j=1}^{N-1} \Big(E^{(f)}_0(j;1) +j(2\mu_1
+4\mu_2) -4\mu_2 \Big)n_j 
\ed
where $E^{(f)}_0(1;1)=0$ by definition. In the limit $N \rightarrow \infty$
the densities $w_j = n_j/N$ must be chosen to minimize
\be\label{e17}
\lim_{N \rightarrow \infty} \frac{E}{N} = \epsilon(w_0,w_1,w_2, \ldots )=
\sum_{j=0}^{+ \infty} \Big(E^{(f)}_0(j;1) +j(2\mu_1 +4\mu_2) -4\mu_2 
\Big)w_j 
\ee
\be\label{e18}
\sum_{j=0}^{+ \infty} (j+1)w_j =1
\ee
As observed in \cite{NUZ}, the problem (\ref{e17}), (\ref{e18}) has arisen in 
classical statistical mechanics of systems with competing 
interactions \cite{FS}. Owing to the linearity of (\ref{e17}) in the 
parameters $\{w_j\}$, the extremum occurs when one $w_j \neq 0$
and all others are zero. Then 
\bd
w_{j} = \frac{1}{j+1} \hspace{2cm} w_{k} = 0 \hspace{1cm}(k \neq j)
\ed
and the ground state energy per site is
\be\label{e19}
\epsilon= \epsilon(j) = \frac{E^{(f)}_0(j;1) +j( 2\mu_1 + 4\mu_2) -
4\mu_2}{j+1}
\ee
The relevant phase is denoted $<j>$.
One is left with the task of determining the minimum in the sequence of real
quantities (\ref{e19}), $j \geq 0$ where it is understood that 
$\epsilon(0)=0$. The minimum
might be reached at $j \rightarrow \infty$, or $\epsilon(\infty) = e_0
+2\mu_1 +4\mu_2$ where $e_0$ is the ground state energy per site of the
spin chain, independent from boundary conditions. The thermodynamical
analysis outlined above does not allow to distinguish between the periodically
closed sector $N^{(1)}=N$ and the open sector where, for instance, one singlet
is introduced. In appendix B it is shown in detail that, when $\inf_{j \geq 0} 
\epsilon(j) = \lim_{j \rightarrow \infty} \epsilon(j)$, then the ground
state indeed belongs to the sector $N^{(1)}=N$.

All said so far is true for any value of $\theta$ in (\ref{e11}). 
Yet, a comparison of
$\epsilon(j)$ requires the knowledge of $E^{(f)}(j;1)$, in principle for
any $j$. The BA solvable cases $(a)$, $(b)$, $(c)$ are perhaps not the most
physically relevant, but they allow an efficient computation of
$E^{(f)}(j;1)$. In the following I will mostly concentrate on case $(c)$, the
purely biquadratic spin 1 chain. The information available for this chain
is summarized in Appendix A. The relevant results on finite size ground state
energies are gathered in Table \ref{tab1}. Some considerations on models
with arbitrary $\theta$ are postponed to section VI.

It can be shown that the $(\mu_1,\mu_2)$ plane is divided into four regions,
corresponding to four different ground state energies per site, hence
four different phases. Since $\mu_1$ and $\mu_2$ often appear in the
combination $2\mu_1+4\mu_2$ define
\bd
\mu_3=2\mu_1+4\mu_2
\ed
First, seek the $(\mu_2,\mu_3)$ values for which $\epsilon(0)$ is lowest,
that is $0 < \epsilon (j)$, $j \geq 1$. It is convenient to rewrite 
this condition 
adding and subtracting a term containing $e_0$ (whose numerical value
is reported in (\ref{a1}))
\bd
4\mu_2 < s_0(j) +j(e_0+\mu_3) \hspace{0.5cm}
j \geq 1 \hspace{2cm} s_0(j) \stackrel{def}{=} E^{(f)}_0(j;1) - je_0 
\ed
The sequence $s_0(j)$ is bounded, so
a first necessary condition is $e_0+\mu_3 > 0$. Furthermore, 
from Table 2, $s_0(j)$ is increasing in the range $2 \leq j
\leq 51$, with a minimum
\be\label{e20}
s_0(2) = E^{(f)}_0(2;1) - 2e_0 = -4 -2e_0 \simeq 1.5937
\ee
What can be said for $j > 51$? From Appendix A, $s_0(j)$ has different limits
at $j \rightarrow \infty$ for $j$ even $(e^{(s,+)})$ and $j$ odd
$(e^{(s,-)})$ but they are both larger than $-4 -2e_0$. It seems, from Table
\ref{tab2}, that $s_0(j)$ at $j \simeq 51$ is already rather close 
to the asymptotic 
regime, so one can safely, although not completely rigorously, conclude
that $s_0(j)$, $j \geq 2$, is minimal at $j=2$ and that, since $e_0 +\mu_3
> 0$
\bd
\inf_{j \geq 2} \Big(s_0(j)+j(e_0+\mu_3)\Big) = s_0(2) +2(e_0+\mu_3) = 
-4 +2\mu_3
\ed
Hence, the three necessary {\em and} sufficient conditions for being in
phase $<0>$ are
\be\label{e21}
\left\{ \begin{array}{lll}
&e_0+\mu_3 > 0  \\
&2\mu_3 -4 \mu_2 > 4  \hspace{1cm} ({\rm phase} <0>) \\
&\mu_3 -4\mu_2 > 0
\end{array} \right.
\ee
The third condition comes from $\epsilon(0)< \epsilon(1)$, considered 
separately.

Phase $<1>$. The conditions for its existence, $\epsilon(1) < \epsilon(0)$
and $\epsilon(1) < \epsilon(j)$, $j \geq 2$ are rewritten, from (\ref{e19})
\barrr
&\mu_3 -4\mu_2 < 0 \hspace{1cm}  \mu_3 +4\mu_2 > s_1(j)  \\
&s_1(j) \stackrel{def}{=} -2E^{(f)}(j;1)/(j-1) \hspace{1cm}  (j \geq 2)
\earrr
where $s_1(j)$ is listed in Table \ref{tab2}, up to $j=51$. 
It is decreasing in this
range and it certainly is at large $j$. In fact, from the known numerical
values (\ref{a1}), (\ref{a7}) and (\ref{a8}), $2e_0+2e^{(s,\pm)}<0$; 
moreover
\bd
s_1(j) = \frac{-2e_0j-2e^{(s,\pm)}}{j-1} +o(1/j) =
-2e_0 - \frac{2e_0+2e^{(s,\pm)}}{j} +o(1/j) \ \ \ (j \rightarrow \infty)
\ed
Hence one can conclude that $\sup_{j \geq 2}s_1(j)=s_1(2)=8$ 
and phase $<1>$ appears when
\be\label{e22}
\mu_3-4\mu_2 <0 \ \ \ \ \mu_3+4\mu_2 > 8 \hspace{2cm} ({\rm phase} <1>)
\ee
Phase $<2>$. The conditions for its existence are $\epsilon(2) <
\epsilon(0)$, $\epsilon(2) < \epsilon(1)$ and $\epsilon(2) < \epsilon(j)$,
$j \geq 3$, or, remembering that $E^{(f)}_0(2;1) = -4$
\barrr
&\mu_3-2\mu_2 < 2 \hspace{1cm} \mu_3+4\mu_2 < 8 \hspace{1cm} \mu_3 +4 \mu_2 >
s_2(j) \hspace{0.5cm} (j \geq 3) \\
&s_2(j) \stackrel{def}{=} \Big(3E^{(f)}_0(j;1)+4(j+1)\Big)/(-j+2) 
\hspace{1cm} (j \geq 3)
\earrr
The sequence $s_2(j)$ is listed in Table \ref{tab2} and it is 
increasing up to $j=51$. It certainly is for large $j$ where
\bd
s_2(j)=\frac{3e_0j+3e^{(s,\pm)}+4(j+1)}{-j+2} +o(1/j) =
-3e_0-4+\frac{-6e_0-3e^{(s,\pm)}-12}{j}+o(1/j)
\ed
because $-6e_0-3e^{(s,\pm)}-12 < 0$. This time, it is not so straightforward
to guess the upper limit of $s_2(j)$. If $s_2(j)$ keeps increasing for
$j>51$ up to the asymptotic regime where one can rely on the previous 
equation, then $\sup_{j \geq 3} s_2(j) =
\lim_{j \rightarrow +\infty} s_{2}(j)= -3e_0-4 \simeq 4.3906$ 
and phase $<2>$ appears for
\be\label{e23}
\mu_3-2\mu_2<2 \hspace{1cm} -3e_0-4 < \mu_3+4\mu_2 < 8 \hspace{1cm}
({\rm phase} <2>) 
\ee
On the other hand, one cannot rule out the possibility that
$\sup_{j \geq 3}s_2(j)$ lie slightly above and this would leave a tiny
window for an additional phase. However, the existence of phase $<2>$ is
unquestionable.

Finally, phase $<\infty>$, whose ground state is that of the periodic
spin 1 chain, shows up when, for each $j$, $\epsilon(j) < \epsilon(\infty)
\stackrel{def}{=} \lim_{j \rightarrow \infty} \epsilon(j)$. Since
$\epsilon(\infty)=e_0+\mu_3$, this means
\barrr
&e_0+\mu_3<0 \hspace{1cm} e_0+\mu_3<(\mu_3-4\mu_2)/2 \\
&e_0+\mu_3 < s_0(j)-4 \mu_2 \hspace{1cm} (j \geq 2)
\earrr
We already know that $s_0(j)$ is minimal at $j=2$, so the third condition 
amounts to $e_0+\mu_3+4\mu_2 < s_0(2) =
-4-2e_0$ which, by itself, implies the second inequality. Hence
\be\label{e24}
e_0+\mu_3<0 \hspace{1cm} \mu_3+4\mu_2 < -4-3e_0 \hspace{1cm}
({\rm phase} <\infty>)
\ee
are the necessary and sufficient conditions that guarantee the appearance 
of phase $<\infty>$. Appendix B completes the proof: the ground state
is that of the periodic spin 1 chain because, under conditions (\ref{e24}), 
local variations around the sector $N^{(1)}=N$ only lead to an increase in
energy. 

Under the foregoing assumption on the upper limit of sequence $s_2(j)$
there is no room for other phases because inequalities 
(\ref{e21})-(\ref{e24}), with
their separating lines, fill the whole $(\mu_2,\mu_3)$ plane.
The ground state energy per site is $\epsilon(0)=0$ (phase $<0>$),
$\epsilon(1)= (\mu_3-4\mu_2)/2$ (phase $<1>$), $\epsilon(2) = 
(-4+2\mu_3-4\mu_2)/3$ (phase $<2>$) and $\epsilon(\infty)=e_0+\mu_3$
(phase $<\infty>$). It shows first derivative discontinuities at the
boundaries.

Here are some general features of the four phases. Define the total spin
\bd
{\bf S}_{{\rm tot}}=\sum_{k=1}^N {\bf S}_k \hspace{2cm}
{\bf S}_{{\rm tot}}^2 = S_{{\rm tot}}(S_{{\rm tot}}+1)
\ed
In phase $<\infty>$ the ground state is simply that of the periodic
biquadratic spin 1 chain. Supposing $N$ even, it is a global singlet and,
it was conjectured in \cite{Kl}, its degeneracy should be 2. Phase $<0>$ has
certainly a unique ground state and of course $S_{{\rm tot}}=0$.
As to phase $<2>$, the ground state
is, again, a tensor product of local singlets, hence a global singlet,
$S_{{\rm tot}}=0$. In fact, it is represented by the sequence
$|\ldots, s,t,t,s,t,t,s,t,t,s, \ldots >$ where neighboring triplets
$|t_k,t_{k+1}>$ are locked into the singlet ground state of 
$-({\bf S}_k \cdot {\bf S}_{k+1})^2$.
Supposing $N=0$ (mod 3) to avoid problems with AFM seams,
the ground state is threefold degenerate: rung singlets sit on rungs 
labelled $k$ (mod 3), with the freedom of choice $k=0,1,2$. 
It clearly spontaneously breaks translational invariance.
Finally, phase $<1>$, described by the sequence 
$|\ldots,t,s,t,s,t,s, \ldots>$
is rather different because nothing fixes the state of each isolated 
triplet resulting in a huge degeneracy, $2\cdot 3^{N/2}$.
Not only is translational invariance spontaneously broken, but 
one is dealing with $N/2$ effectively noninteracting triplets that can
combine in a ``ferrimagnetic'' state $S_{{\rm tot}}=N/2$, or combine 
in local pairs yielding $S_{{\rm tot}}=0$ (these are only two of 
the many possibilities). In other words, the ground state is not necessarily
an eigenstate of ${\bf S}_{{\rm tot}}$.

\section{Excitations}

The four phases display a rich variety of excitations. Most of them,
especially in the fragmented phases $<0>$, $<1>$ and $<2>$ can easily be
determined resorting to (\ref{e16}), therefore only phase $<\infty>$ will be
extensively treated in the following. Appendix B gives a complete
discussion of how, in phase $<\infty>$, ``local'' perturbations of
the biquadratic ground state can only lead to an increase of energy.

A first kind of excitations, within the sector $N^{(1)}=N$, are those of
the biquadratic chain itself. Supposing $N$ even, they exsist in 
even number and have a gap (see Appendix A for a more 
exhaustive discussion)
\bd
\Delta E_{{\rm gap}} \simeq 0.173179
\ed

A second kind of excitation, called TSST (triplet-to-singlet spin flip)
in \cite{Xi}, is obtained by introducing one singlet, 
sector $N^{(1)}=N-1$. There is $N$ ways to do so and the eigenvalue is
$N$-fold degenerate. This implies that linear superpositions of such states
produce eigenstates of the shift operator, all with the same energy.
As a matter of fact, dispersionless excitations have already been found
in akin ladders \cite{KM}. The energy difference is, see (\ref{e16})
\barrr
&&\Delta E(N) = E^{(f)}_0 (N-1;1)-E^{(p)}_0(N;1)-\mu_3-4\mu_2 \\
&&\Delta E = \lim_{N \rightarrow \infty} \Delta E(N) =
e^{(s,-)} - e_0 -\mu_3-4\mu_2 
\earrr
From conditions (\ref{e24}) it follows that 
$\Delta E > 2e_0 +4 +e^{(s,-)} \simeq 0.1479$ confirming that 
the energy can only be increased by fragmentation.
At the same time, that numerical value shows that, at least near the boundary
lines, the energy scale is comparable, actually slightly smaller, than
$\Delta E_{{\rm gap}}$. Therefore fragmentation may be relevant 
to the low energy physics of the ladder.
Despite what might seem, such excitations depend on two degrees of freedom. 
In fact, as discussed in Appendix A, the ground state of the fragment of 
triplets, i.e. the ground state of the open biquadratic chain with
$N-1$ (odd) sites, is only the lowest in a continuous band. This adds a
second degree of freedom to the singlet's position. Therefore, if $n$
is the singlet position
\bd
\Delta E(\alpha,n) = \epsilon(\alpha) - \epsilon(\pi) + e^{(s,-)}-e_0
-\mu_3 -4\mu_2 \hspace{2cm} \alpha \in (0,\pi)
\ed
Here $\epsilon(\alpha)-\epsilon(\pi)$, always positive, measures the
excitation energy of the single kink in the open spin chain and vanishes
at $\alpha \rightarrow \pi$, see (\ref{a5}). The $n$-dependence is, of course,
trivial. Nonetheless, it is at least plausible that under a small variation
of coupling constants in (\ref{e1}), for instance a different relative weight
between $I^{(1)}$ and $I^{(3)}$, or $I^{(4)}$ and $I^{(6)}$ interactions,
$\Delta E$ might acquire a genuine two-parameter form.
Since the triplets fragment is presumably
locked into a spin 1 state (see Appendix A), $S_{{\rm tot}}=1$ for 
such states.
Triplet-singlet excitations have been found in similar ladders, but 
they arise through a different mechanism \cite{NT,KM}.

When two rung singlets are introduced ($N^{(1)}=N-2$), their energy becomes
position dependent. If they sit on rungs $n_1,n_2$ and 
$d=n_2-n_1 \geq 1$, then from (\ref{e16})
\bd
\Delta E(N;n_1,n_2)=E^{(f)}_0(d-1;1)+E^{(f)}_0(N-d-1;1)-E^{(p)}_0(N;1)-
2\mu_3-8\mu_2 +4\mu_2 \delta_{d,1}
\ed
If $N \rightarrow \infty$, keeping $n_1$, $n_2$ fixed ($p(d)$ is the parity
of $d$)
\be\label{e25}
\Delta E(d) = \left\{ \begin{array}{lr}
s_0(d-1)-2e_0+e^{(s,p(d-1))}-2\mu_3-8\mu_2 \hspace{1cm} & (d > 2) \nonumber \\
-3e_0 + e^{(s,-)} -2\mu_3 -8\mu_2   & (d=2) \\
-2e_0 + e^{(s,+)} - 2\mu_3 -4\mu_2  & (d=1)  \nonumber
\end{array} \right.
\ee
Notice that all these excitation energies are exact.
It is easy to see that $\Delta E(d) > 0$ always. It represents a sort of
``static interaction energy'' between singlets, whose distance  dependence
is governed by $s_0(d)$. The open chain surface energy is different for
even and odd lengths, so $s_0(d)$ has a uniform $d$-dependence only if
the parity of $d$ is fixed. When $d$ is even, $\Delta E(d)$ uniformly 
decreases (see Table \ref{tab2}) from a maximum $\Delta E(2)$ to a minimum 
$\lim_{d \rightarrow
\infty} \Delta E(d) = -2e_0 +2e^{(s,-)} -2\mu_3 -8\mu_2$.
When $d$ is odd, the behavior is certainly uniform for $d \geq 3$, because
$\Delta E(d)$ increases from $\Delta E(3)$ to a maximum $\lim_{d \rightarrow
\infty} \Delta E(d) = 2e^{(s,+)}-2e_0-2\mu_3-8\mu_2$. Still, $\Delta E(1)$
is not necessarily lower or higher than $\Delta E(3)$, since
\bd
\Delta E(3)-\Delta E(1) = -4 -2e_0 -4\mu_2
\ed
whose sign is left undetermined from conditions (\ref{e24}). The short range
behavior can be either repulsive or attractive.

A third possible kind of excitations, named $n$-TSSF in \cite{Xi},
is produced by the insertion of a block of $m$ singlets, represented 
by $|\ldots, t,t,t,s,s,s,s,t,t,t,t, \ldots >$. Now, from (\ref{e16}), 
supposing $N$ even as always
\barrr
&&\Delta E(N;m) = E_0^{(f)}(N-m;1)- m\mu_3 -4\mu_2 -E_0^{(p)}(N;1) \\
&&\Delta E(m) =\lim_{N \rightarrow \infty} \Delta E(N;m) =
-(m-1)(e_0+\mu_3) + e^{(s,p(m))} -e_0-\mu_3-4\mu_2
\earrr
which is positive under conditions (\ref{e24}). At $e_0+\mu_3 =0$ these
perturbations determine the instability versus phase $<0>$.

\section{Other values of $\theta$}

It is easy to see that phases $<0>$, $<1>$ and $<\infty>$ will be
present no matter what value $\theta$ takes in (\ref{e11}). In this
general situation, (\ref{e16}) will formally be the same but now
with $E_0^{(f)}(N_j^{(1)};1;\theta)$ and $E_0^{(p)}(N;1;\theta)$,
the energies of the spin 1 chain (\ref{e3}) with arbitrary $\theta$.
The proof of section 4 goes through unchanged. One has to find the
minimum of
\bd
\epsilon(j;\theta)= \frac{E_0^{(f)}(j;1;\theta)+j\mu_3-4\mu_2}
{j+1} \hspace{1cm} \epsilon(0;\theta) = 0
\ed
The conditions for phase $<0>$ amount to
\be\label{ad1}
4\mu_2 < s_0(j)+j(e_0(\theta)+\mu_3) \hspace{0.5cm} (j \geq 1) \hspace{1cm}
s_0(j) \stackrel{def}{=} E_0^{(f)}(j;1;\theta)-e_0(\theta)j
\ee
Notice that $s_0(j)$ is bounded. Set
\bd
s_0(j_1)= \inf_{j \geq 1} s_0(j)
\ed
Of course $j_1$ might be $+ \infty$. In this case, since $\lim_{j
\rightarrow \infty}s_0(j) = e^{(s,\pm)}$, the two possible surface energies,
one has $s_0(j_1) = {\rm min}\{e^{(s,+)},e^{(s,-)}\}$. 
At any rate, $s_0(j_1)$ is finite. A
necessary condition required by (\ref{ad1}) is $e_0(\theta) +\mu_3 >0$,
but since one does not know the exact value $j_1$ one can only give
sufficient conditions for the existence of $<0>$, namely
\barrr
&&e_0(\theta)+\mu_3 > 0  \hspace{2cm}  ({\rm phase} <0>)  \\
&&4\mu_2 < \inf_{j \geq 1}s_0(j) + \inf_{j \geq 1} j(e_0(\theta)+\mu_3) =
s_0(j_1) + e_0(\theta) +\mu_3
\earrr
which can always be simultaneously satisfied by suitable values 
of $(\mu_2,\mu_3)$.

Likewise, for phase $<\infty>$, the condition $\epsilon(j) > e_0(\theta)
+ \mu_3$, $(j \geq 0)$, translates into $e_0(\theta) +\mu_3 < 0$ and
$e_0(\theta) +\mu_3 +4\mu_2 < s_0(j)$, $(j \geq 1)$, which can both be
fulfilled taking
\bd
e_0(\theta) +\mu_3 < 0 \hspace{1cm} e_0(\theta)+\mu_3+4\mu_2 < s_0(j_1)
\hspace{2cm}  ({\rm phase} <\infty>)
\ed
These conditions are necessary and sufficient, and are compatible.

Finally, the inequalities for $<1>$ are rewitten
\bd
\mu_3 -4\mu_2 < 0 \hspace{2cm} \mu_3+4\mu_2 >- \frac{2 E_0^{(f)}(j;1;\theta)}
{j-1} \stackrel{def}{=} s_1(j) \hspace{0.5cm} j \geq 2
\ed
Sequence $s_1(j)$ is also bounded. Set $s_1(j_2) = \sup_{j \geq 2}
s_1(j)$. The ensuing necessary and sufficient conditions are
\bd
\mu_3-4\mu_2 < 0 \hspace{1cm} \mu_3+4\mu_2 > s_1(j_2) \hspace{2cm}
({\rm phase} <1>)
\ed
On the contrary, it is in general impossible to draw conclusions on the
existence of other phases. For example, phase $<2>$ would be present 
if it were possible to fulfill simultaneously
\be\label{ad2}
4\mu_2-2\mu_3 > E_0^{(f)}(2;1;\theta) \hspace{1cm} \mu_3+4\mu_2 <
-2E_0^{(f)}(2;1;\theta)  \hspace{1cm} \mu_3+4\mu_2 > s_2(j) 
\ee
\bd
s_2(j) \stackrel{def}{=} \Big((j+1)E_0^{(f)}(2;1;\theta)-
3E_0^{(f)}(j;1;\theta) \Big)/(j-2) \hspace{1cm} (j \geq 3) 
\ed
Notice that $s_2(j)$ is also bounded. If we set $s_2(j_3)= \sup_{j \geq 3}
s_2(j)$, the second and the third inequality in (\ref{ad2}) are compatible 
if and only if
\be\label{e27'}
s_2(j_2) < -2E_0^{(f)}(2;1;\theta)
\ee
At least for $\theta =0$, $e_0(\theta)$ has been determined to a great
accuracy and one can attempt to estimate whether (\ref{e27'}) holds.
It is known that \cite{WH}
\bd
e_0(\theta=0) \simeq -1.401484
\ed
On the other hand, not many data seem to have been published for
$E_0^{(f)}(j;1;0)$ even at small $j$. Diagonalization of (\ref{e3}),
with free boundaries, can easily be carried out numerically 
up to 7 sites (as a test, I diagonalized the periodic chain as well and
compared the results with those published in \cite{Bl}). It is
sufficient to examine the sector $S^z_{tot}=0$ because, owing to 
$SU(2)$-invariance, all possible eigenvalues are bound to appear there.
Now, $E_0^{(f)}(2;1;0)=-2$, while the sequence $\{s_2(j)\}_{j=3}^7$ is
found to have the approximate values $\{1,1.9606,1.8302,2.0277,1.9807\}$.
Furthermore
\bd
\lim_{j \rightarrow \infty} s_2(j) = -2 -3e_0 \simeq 2.20446
\ed
It is of course impossible to draw a rigorous conclusion from these results,
but it looks extremely likely that inequality (\ref{e27'}) is true and
therefore phase $<2>$ is present also for $\theta =0$.

\section{Discussion}

It is useful to compare the ladders considered here with those studied
in previous papers. To shorten notation, I will denote by 
$|\underline{0}>_{<x>}$
the ground state of phase $<x>$, or one of them if it is degenerate.
In the work of Niggeman {\em et al.} similar methods were employed but
for different systems, namely the mixed Heisenberg chains. Perhaps the
most closely related ladder is that of Xian \cite{Xi}, which is (\ref{e11}) 
for $\theta =0$ and $\mu_2 =0$. Perturbations of this ladder where 
considered in \cite{HHR,WKO}. Xian finds only two phases, 
$<0>$ and $<\infty>$ in the present notation, but, since $\theta=0$, his 
ground state $|\underline{0}>_{<\infty>}$ is that
of the spin 1 Heisenberg chain. The absence of phases $<1>$ and $<2>$
can be understood when one notices that even for the case considered
in this paper, $\theta = - \pi/2$, phases $<1>$ and $<2>$ would be missing
without the rung-rung interaction $I^{(5)}_{k,k+1}$, as can be 
seen by setting $\mu_2 =0$ in (\ref{e22}) and (\ref{e23}).

The ground state of a wide class of $SU(2)$-invariant ladders has the
matrix-product (MP) form \cite{KM}. Such states can be 
translationally invariant or dimerized
\barrr
|\psi_0^{({\rm inv})}(u)>&=&{\rm Tr} \prod_{n=1}^N g_n(u) \hspace{1cm}
|\psi_0^{({\rm dim})}(u_1,u_2)> = {\rm Tr} \prod_{n=1}^{N/2}g_{2n-1}(u_1)
g_{2n}(u_2)  \\
g_n(u) &=& \left[ \begin{array}{cc}
u |s>_n + |0>_n & - \sqrt{2}|1>_n \\
\sqrt{2} |-1>_{n} & u|s>_n - |0>_n
\end{array} \right]
\earrr 
For models with translationally invariant MP ground states, 
$u=0$ and $u=\infty$ are the only points which have a significant overlapping
with the ladders presented here. At $u=0$ the MP state is made up of rung
triplets and is, effectively, the ground state of the AKLT spin 1 chain.
That is what one gets for (\ref{e11}) by setting 
$\tan \theta =1/3$ and keeping $\mu_1$, $\mu_2$ small enough. 
Now, $|\underline{0}>_{\infty}$ 
is also made up of triplets, but it is that of the biquadratic 
spin 1 chain, hence a different kind. Furthermore, 
$|\underline{0}>_{<0>}$, $|\underline{0}>_{<1>}$ and 
$|\underline{0}>_{<2>}$ all contain rung singlets and are 
certainly not $|\psi_0^{({\rm inv})}(0)>$. Conversely,
after suitable normalization, the limit $|\psi_0^{({\rm inv})}(\infty)>$
is made up entirely of rung singlets. There is only one such state 
so it must coincide with $|\underline{0}>_{<0>}$. Of course excitations 
may differ according to the detailed form of the hamiltonian. 
At any rate, $|\underline{0}>_{<1>}$ and $|\underline{0}>_{<2>}$ 
break translational invariance so they can never be of
the form $|\psi_0^{({\rm inv})}(u)>$ for any $u$. For the same reason,
$|\underline{0}>_{<2>}$, which is invariant under 
a three-rung translation, can never be of the form 
$|\psi_0^{({\rm dim})}(u_1,u_2)>$ either, since the latter is 
invariant under a two-rung translation.
But, within the large lowest energy eigenspace of phase $<1>$, at least
one vector is in MP form. To show it, notice that at finite, 
non-vanishing values of $u_1$, $u_2$, 
$|\psi_0^{({\rm dim})}(u_1,u_2)>$ is a linear combination necessarily 
containig kets with all triplets and the ket with all singlets. The only
way to reproduce $|\underline{0}>_{<1>}$ is to set $u_1=0$ 
and to take $u_2 \rightarrow 
\infty$ after proper normalization of $g_{2n}(u_2)$. The outcome is a
linear combination of kets which have singlets on even rungs and triplets on
odd rungs, exactly like $|\underline{0}>_{<1>}$. It should be 
observed though the the MP approach only in few cases allow to 
find the whole spectrum, as it is instead possible from BA for
hamiltonian (\ref{e1}).

\section*{Acknowledgements}
I am grateful to M.Raciti and F.Riva for useful discussions.

\appendix

\section{The biquadratic spin 1 chain}

The purely biquadratic spin 1 chain is integrable for periodic 
\cite{Kl} and free \cite{BB,BH} boundary conditions. The definition 
of the relevant elliptic parameters differs in the two cases, 
as treated in the literature. For the sake of readability,
a Landen transformation \cite{Bax} has been applied
to the results of \cite{Kl}, in order to make notations compatible.

Define an elliptic modulus from the relation
\bd
\frac{K'(k)}{K(k)} = \frac{1}{\pi} \ln(1/q) \hspace{2cm} 
q=\frac{3-\sqrt{5}}{2}
\ed
where $K(k)$ and $K'(k)$ are the elliptic integrals of first and second kind.
The ground state energy, which is the same in the two cases, is then
\be\label{a1}
e_0=-1-\frac{\sqrt{5}}{2} \Big(1+4 \sum_{n=1}^{+\infty} \frac{q^{2n}}
{1+q^{2n}} \Big) \simeq -2.796863...
\ee
For the periodic case, energy and momentum of the excitations are given,
in the limit $N \rightarrow \infty$, by
\barr\label{a2}
&& \Delta E = \sum_{i=1}^{2\nu} \tilde{\epsilon}(p_i) \hspace{2cm}
\Delta P = \sum_{i=1}^{2\nu}p_i  \nonumber \\
&&\tilde{\epsilon}(p) = \frac{\sqrt{5}}{\pi} K(k) \sqrt{k'^{2} + k^2 \sin p}
\hspace{2cm} k'^2 + k^2 =1
\earr
There is a gap in the spectrum
\be\label{a3}
\Delta E_{{\rm gap}}= 2 \tilde{\epsilon}(0) =2\frac{\sqrt{5}}{\pi}
K(k) k' = \sqrt{5} \prod_{n=1}^{+ \infty} \Big(\frac{1-q^n}{1+q^n})^2
\simeq 0.173178...
\ee
The free boundary case has been solved by noticing that the spectrum is
the same, up to degeneracies, as that of the $XXZ$ spin 1/2 chain with a
boundary field \cite{BB}
\barr
&&H_{XXZ} = -\frac{1}{2} \sum_{k=1}^{N-1}(\sigma_{k}^x \sigma_{k+1}^x +
\sigma_{k}^y \sigma_{k+1}^y - \cosh \gamma \sigma_{k}^z \sigma_{k+1}^z )
+ \frac{\sinh \gamma}{2}(\sigma_1^z -\sigma_{N}^z) \label{a4} \\
&& E_{{\rm bQ}} = E_{XXZ} - \frac{7}{4}(N-1) 
\hspace{2cm} q=e^{-\gamma} \nonumber
\earr
The mapping, valid for any $N$, works because both hamiltonians can be
written as sums over generators of the same Temperley-Lieb algebra \cite{BB}.
Consequently, ground state energies of (\ref{a4}) are for any $N$ identical,
up to a shift, to ground state energies of the for biquadratic chain with
free boundaries. In turn, (\ref{a4}) has been solved by coordinate
Bethe-ansatz \cite{ABBBQ}, so eigenvalues are found from solutions 
of a set of coupled trascendental equations
\barr\label{a4'}
&&\frac{1}{\pi}\Theta(\alpha_j;\frac{\gamma}{2})- \frac{1}{2\pi N}
\sum_{k=1, k \neq j}^{n} \Big( \Theta(\alpha_j- \alpha_k ; \gamma) + \Theta
(\alpha_j+ \alpha_k ; \gamma) \Big) = \frac{I_j}{N} \hspace{1cm} 
j=1 \ldots n \hspace{1cm} \nonumber \\
&&\Theta(\alpha ; x) \stackrel{def}{=} -i\ln \left[ \frac{\sinh(x+
\frac{i\alpha}{2})}{\sinh(x-\frac{i\alpha}{2})} \right] 
= 2\arctan(\tan \frac{\alpha}{2} \coth x)   \\
&&E=\frac{1}{2}(N-1)\cosh \gamma -2 \sinh \gamma \sum_{j=1}^{n} \Theta' 
(\alpha_j ;\frac{\gamma}{2}) \nonumber
\earr
The function $\Theta(\alpha;x)$ is defined to be continuous for real 
$\alpha$. The $\{I_j\}$ in (\ref{a4'}) are positive integers
\cite{ABBBQ,BH}. The ground state belongs to the sector $n=N/2$ ($N$ even) or
$n=(N-1)/2$ ($N$ odd). In both cases, in (\ref{a4'}), $I_j=j$, 
$j=1,2 \ldots n$. Data in Table \ref{tab1} have been obtained 
by numerically solving (\ref{a4'}) with this choice of integers.

Excitations are, in general, gapful (no dispersion relation can be written
here because linear momentum is not conserved).
\barrr
&& \Delta E = \sum_{i=1}^{2\nu}\epsilon(\alpha_i) \hspace{2cm}
\epsilon(\alpha) = 2\sinh \gamma \frac{K(k)}{\pi} {\rm dn}(\frac{K(k)
\alpha}{\pi};k) \hspace{1cm} \alpha \in (0,\pi) \\
&& \Delta E_{\rm gap} = 2\epsilon (\pi) = \sqrt{5} \prod_{n=1}^{+ \infty}
\Big(\frac{1-q^n}{1+q^n}\Big)^2
\earrr
as in (\ref{a3}). When $N$ is odd, though,the ground state is at the 
bottom of a one-paramenter, continuous, gapless (in the 
$N \rightarrow \infty$ limit) band of eigenvalues \cite{Al}. 
Within this band
\be\label{a5}
\Delta E (\alpha) = \epsilon(\alpha) -\epsilon(\pi) \hspace{2cm}  
\alpha \in (0,\pi)
\ee
This may be interpreted by saying that the ground state at $N$ odd is
not the ``vacuum'', but it contains one particle (kink) which can take
up a whole band of dynamical states.

Surface energies for open antiferromagnetic chains differ, in general,
when the limit $N \rightarrow \infty$ is taken for $N$ even $(e^{(s,+)})$
or $N$ odd $(e^{(s,-)})$. They are defined by
\be\label{a6}
E_0(N) = e_0N + e^{(s,\pm)} + o(1) \hspace{1cm} (N \rightarrow \infty)
\ee
and, for the case at hand, they can be computed exactly from (\ref{a4'}). 
For the biquadratic chain, one finds \cite{BH,Al}
\barr
&&e^{(s,+)} = 1 +2\sqrt{5} \sum_{n=1}^{+ \infty} 
\frac{q^{2n}-q^{4n}}{1+q^{4n}} \simeq 1.6550092  \label{a7} \\
&&e^{(s,-)} = 1 + \frac{\sqrt{5}}{2} \Big(1+4 \sum_{n=1}^{+ \infty}
\frac{q^{2n}-q^{4n}}{1+q^{4n}}- \sum_{n=0}^{+ \infty} \frac{4q^{2n+1}}
{1+q^{4n+2}} \Big) \simeq 1.7415986 \label{a8}
\earr

\section{Ground state proof for phase $<\infty>$}

It has to be proven that, within the boundaries of region (\ref{e24}), 
the ladder ground state indeed belongs to the sector $N^{(1)}=N$, 
therefore it is just the ground state of (\ref{e3}) at $\theta = - \pi/2$.
Here it will be shown that it is the lowest energy
state within the class which encompasses the sectors $\{N_j^{(1)}, 
N_j^{(0)} \}_{j=1}^n$ where $\{N_j^{(0)}\}_{j=1}^n$ and 
$\{N^{(1)}\}_{j=1}^{n-1}$
are arbitrarily large but kept finite as $N \rightarrow \infty$. In
other words, only one triplet fragment, $N^{(1)}_n$ has a diverging size
in the thermodynamic limit, as dictated by the analysis of section 4. 
It will be assumed that $\inf_{j \geq 2} s_0(j) = s_0(2)$.

From (\ref{e16})
\bd
\Delta E(N)=\sum_{j=1}^{n-1}E_0^{(f)}(N_j^{(1)};1) + E_0^{(f)}(N_n^{(1)};1)
+\mu_3 N^{(1)} -4n\mu_2 - (E_0^{(p)}(N;1)+\mu_3 N)
\ed
When $N \rightarrow \infty$, setting $\Delta E = \lim_{N \rightarrow \infty}
\Delta E(N)$
\bd
\Delta E= \sum_{j=1}^{n-1}E_0^{(f)}(N_j^{(1)};1) +e_0 N_j^{(1)}
+e^{(s,p(N_n^{(1)}))}- \mu_3 N^{(0)} -4n \mu_2 -e_0 N 
\ed
Since $N^{(1)}_n=N-N^{(0)}-\sum_{j=1}^{n-1} N_j^{(1)}$ and $e^{(s)} \geq
e^{(s,+)}$
\be\label{b1}
\Delta E \geq \sum_{j=1}^{n-1} \Big( E_0^{(f)}(N_j^{(1)};1)-e_0 N_j^{(1)} \Big)
-(e_0 + \mu_3 )N^{(0)} +e^{(s,+)} -4n \mu_2 
\ee
The bracketed term is $s_0(N_j^{(1)})$. Actually $s_0(j)$ has been defined
for $j \geq 2$, whereas in (\ref{b1}) $N_j^{(1)}$ can be 1 for some $j$. The
definition of $s_0(j)$, though, makes sense for $j=1$, too. In that case
$s_0(1) =-e_0 > s_0(2)$, so it is also true that $\inf_{j \geq 1}
s_0(j) = s_0(2)$. Then
\bd
\Delta E \geq (n-1)s_0(2) + e^{(s,+)} - (e_0 + \mu_3)N^{(0)} -4n \mu_2 >
n s_0(2) - (e_0+ \mu_3)N^{(0)} -4n \mu_2
\ed
because $e^{(s,+)} > s_0(2)$. Next, notice that
\bd
N^{(0)} \geq n
\ed
because to create a new fragment at least one singlet must be added. So,
setting $N^{(0)} = n + \Delta N^{(0)}$,
\bd
\Delta E > -( e_0 + \mu_3) \Delta N^{(0)} + n(s_0(2) - e_0 - \mu_3-4
\mu_2)
\ed
which is positive due to inequalities (\ref{e24}).

\newpage
\pagestyle{empty}
\narrowtext
\begin{table}
\caption{Ground state energies of the biquadratic chain}
\label{tab1}
\begin{tabular}{ll}
$N$ & $E^{(f)}_0(N;1)$ \\
\tableline
1 & 0 \\ 
2 & -4 \\
3 & -6 \\
4 & -9.56155281 \\
5 & -11.76372382 \\ 
6 & -15.14325669 \\ 
7 & -17.45402100 \\
8 & -20.73101075 \\ 
9 & -23.11096357 \\ 
10 & -26.32131467 \\
11 & -28.74967412 \\
12 & -31.91290040 \\
13 & -34.37723072 \\
14 & -37.50520573 \\
15 & -39.99742180 \\
16 & -43.09794702 \\
17 & -45.61247178 \\
18 & -48.69096768 \\
19 & -51.22377887 \\
20 & -54.28417520 \\
21 & -56.83226912 \\
22 & -59.87751203 \\
23 & -62.43858209 \\
24 & -65.47094086 \\
25 & -68.04317437 \\
26 & -71.06443665 \\
27 & -73.64638147 \\
28 & -76.65798212 \\
29 & -79.24845576 \\
30 & -82.25156507 \\
31 & -84.84959092 \\
32 & -87.84517669 \\
33 & -90.44993829 \\
34 & -93.43881050 \\
35 & -96.04961790 \\
36 & -99.03246169 \\
37 & -101.64872625 \\
38 & -104.62612662 \\
39 & -107.24734191 \\
40 & -110.21980250 \\
41 & -112.84552974 \\
42 & -115.81348717 \\
43 & -118.44334259 \\
44 & -121.40717895 \\
45 & -124.04082624 \\
46 & -127.00087653 \\
47 & -129.63801837 \\
48 & -132.59457884 \\
49 & -135.23495129 \\ 
50 & -138.18828505 \\ 
51 & -140.83165264 \\
\end{tabular}
\end{table}

\newpage
\pagestyle{empty}
\narrowtext
\begin{table}
\caption{Sequences $s_0(n)$, $s_1(n)$, $s_2(n)$ up to $n=51$}
\label{tab2}
\begin{tabular}{llll}
$n$ & $s_0(n)$ & $s_1(n)$ & $s_2(n)$ \\
\tableline
1 & {\rm n.d} & {\rm n.d.} & {\rm n.d.} \\ 
2 & 1.59372686 & 8 & {\rm n.d.} \\ 
3 & 2.39059029 & 6 & 2 \\ 
4 & 1.62590091 & 6.37436854 & 4.34232922 \\ 
5 & 2.22059333 & 5.88186191 & 3.76372382 \\ 
6 & 1.63792389 & 6.05730268 & 4.35744252 \\ 
7 & 2.12402301 & 5.81800700 & 4.07241260 \\ 
8 & 1.64389669 & 5.92314593 & 4.36550538 \\ 
9 & 2.06080730 & 5.77774089 & 4.19041296 \\ 
10 & 1.64731963 & 5.84918104 & 4.37049300 \\ 
11 & 2.01582361 & 5.74993482 & 4.24989137 \\ 
12 & 1.64946076 & 5.80234553 & 4.37387012 \\ 
13 & 1.98199387 & 5.72953845 & 4.28469929 \\ 
14 & 1.65088229 & 5.77003165 & 4.37630143 \\ 
15 & 1.95552965 & 5.71391740 & 4.30709734 \\ 
16 & 1.65186786 & 5.74639294 & 4.37813150 \\ 
17 & 1.93420653 & 5.70155897 & 4.32249436 \\ 
18 & 1.65257406 & 5.72834914 & 4.37955644 \\ 
19 & 1.91662630 & 5.69153099 & 4.33360804 \\ 
20 & 1.65309340 & 5.71412371 & 4.38069587 \\ 
21 & 1.90186291 & 5.68322691 & 4.34193723 \\ 
22 & 1.65348343 & 5.70262019 & 4.38162680 \\ 
23 & 1.88927680 & 5.67623474 & 4.34836887 \\ 
24 & 1.65378146 & 5.69312529 & 4.38240103 \\ 
25 & 1.87841138 & 5.67026453 & 4.35345753 \\ 
26 & 1.65401253 & 5.68515493 & 4.38305458 \\ 
27 & 1.86893114 & 5.66510627 & 4.35756578 \\ 
28 & 1.65419392 & 5.67836905 & 4.38361332 \\ 
29 & 1.86058371 & 5.66060398 & 4.36093953 \\ 
30 & 1.65433783 & 5.67252173 & 4.38409626 \\
31 & 1.85317541 & 5.65663939 & 4.36375078 \\ 
32 & 1.65445307 & 5.66743075 & 4.38451767 \\ 
33 & 1.84655490 & 5.65312114 & 4.36612306 \\ 
34 & 1.65454612 & 5.66295821 & 4.38488848 \\ 
35 & 1.84060215 & 5.64997752 & 4.36814708 \\ 
36 & 1.65462179 & 5.65899781 & 4.38521721 \\ 
37 & 1.83522066 & 5.64715146 & 4.36989082 \\ 
38 & 1.65468372 & 5.65546630 & 4.38551055 \\ 
39 & 1.83033186 & 5.64459694 & 4.37140610 \\
40 & 1.65473470 & 5.65229756 & 4.38577388 \\ 
41 & 1.82587089 & 5.64227649 & 4.37273306 \\ 
42 & 1.65477689 & 5.64943840 & 4.38601154 \\ 
43 & 1.82178490 & 5.64015917 & 4.37390312 \\ 
44 & 1.65481197 & 5.64684553 & 4.38622707 \\ 
45 & 1.81802811 & 5.63821937 & 4.37494137 \\ 
46 & 1.65484125 & 5.64448340 & 4.38642340 \\ 
47 & 1.81456284 & 5.63643558 & 4.37586789 \\ 
48 & 1.65486580 & 5.64232250 & 4.38660297 \\ 
49 & 1.81135678 & 5.63478964 & 4.37669902 \\
50 & 1.65488645 & 5.64033817 & 4.38676782 \\ 
51 & 1.80838229 & 5.63326611 & 4.37744812 \\
\end{tabular}
\end{table}

\end{document}